# Non-equilibrium pathway to mesoscale ordering in ethanol-water binary liquid

Xinyue Jiang[1], Yating Shang[2], Jianhui Li[3], Zhaoyong Zou[3],Yanxia Zuo[4], Yuqun Xie[2*]

[1]School of Civil Engineering Architecture and Environment, Hubei University of Technology, Wuhan 430068, China

[2]School of Life and Health Sciences, Hubei University of Technology, Wuhan 430068, China

[3]State Key Laboratory of Advanced Technology for Materials Synthesis and Processing, Wuhan University of Technology, Wuhan, 430070, China

[4]The Analysis and Testing Center of Institute of Hydrobiology, Chinese Academy of Sciences, Wuhan 43007, China

## Abstract

Ethanol-water mixtures are a classic example of thermodynamic non-ideality, yet the structural origin of their pronounced anomalies, such as volume contraction and a large negative excess entropy, has remained a long-standing puzzle. Here, we demonstrate these anomalies are not equilibrium properties but calorimetric fingerprint of an arrested phase transition. By imposing periodic thermal oscillations, we drive a 50% (v/v) ethanol-water system along a complete hierarchical self-assembly pathway that progressed from ethanol clusters to water-containing droplets, then to acicular flakes, and finally to micron-scale ordered ethanol aggregates. Fluorescence spectroscopy, two-dimensional correlation analysis and nuclear magnetic resonance revealed the underlying non-equilibrium molecular mechanism: a periodic perturbation of the water-dominated hydrogen-bond network initiates a ethanol-water coexistence intermediate, ultimately leading to the stable ordered assembly of an ethanol-rich phase. Our finding demonstrated that periodic physical perturbations capable drive spontaneous ordering across multiple length scales in a simple binary mixture, providing a kinetic perspective on the structural origin of solution non-ideality, and carry general implications for self-assembly strategies in soft matter.

## Main

The ethanol-water system has long been known for its pronounced non-ideality within the equilibrium thermodynamic framework. Classical excess functions precisely quantify its mixing anomalies, including volume contraction, heat release and a large negative excess entropy, with a characteristic inflection point near 50% (v/v)[1, 2, 3, 4]. Our previous work demonstrated that this inflection point correlates with the predominance of ethanol tetramer clusters[5]. Existing studies have largely focused on equilibrium structural characterization using modern techniques[6, 7]. Neutron diffraction revealed concentration-dependent microscopic heterogeneity[8, 9]; dielectric relaxation and terahertz spectroscopy captured multi-timescale hydrogen-bond dynamics[10, 11, 12]; molecular simulations suggested the existence of percolation, cyclic hydrogen-bond motifs and mesoscopic structures of tens to hundreds of nanometres[13, 14, 15, 16, 17]. However, these approaches, which are largely confined to characterizing static structures at equilibrium, inherently fail to capture the dynamic processes that could give rise to them, leaving the structural origin of the anomalies unresolved.

We repeatedly observed mesoscopic flake-like aggregates in the ethanol-water solution, while the temperature sensitivity of the excess thermodynamic functions previously revealed by Franks and Ives provided a key clue for understanding this counter-equilibrium phenomenon[18]. The balance between mixing enthalpy and excess entropy changes sharply with temperature. In certain concentration regimes, the mixing process even switches from exothermic to endothermic. This observation strongly suggest that the stability of ordered structures within the system is governed by temperature, and that temperature variation itself may act as a dynamic driving force to direct molecular assembly pathways. However, previous studies remained within the equilibrium or quasi-equilibrium framework, limited to comparing structural differences at different static temperatures. They overlooked the fact that these thermodynamic anomalies inherently involve a non-equilibrium dynamic process.

Here we introduced periodic thermal oscillation as a controlled non-equilibrium perturbation and directly track the entire evolution path from molecular association to macroscopic order in a 50% (v/v) ethanol-water mixture. Driven by thermal oscillation, this simple binary mixture underwent a clear hierarchical self-assembly process. It progressed from initial ethanol clusters to water-containing droplets, a typical liquid-liquid microphase separation intermediate, then to acicular flakes, and finally stacked into stable ordered ethanol aggregates (Fig. 1). This evolution accompanied a gradual enhancement of hydrophobic domains, a marked increase in the thermal stability of the aggregates, a persistent decrease in free ethanol concentration and a cooperative reorganisation of the hydrogen-bond network in the bulk phase. Through morphology characterisation, thermal analysis, in situ temperature-dependent Raman spectroscopy, gas chromatography-mass spectrometry (GC-MS), fluorescence spectroscopy and nuclear magnetic resonance relaxometry, we constructed a multi-scale evidence chain for this non-equilibrium phase transition path. The results demonstrated that thermal oscillation served not merely to accelerate the system towards equilibrium but to guide it along a specific kinetic pathway. A pathway that started with periodic perturbation and relaxation of the water-dominated hydrogen-bond network, proceeds through a coexisting ethanol-water intermediate stage, and culminates in the stable ordered assembly of an ethanol-rich phase.

Our findings establish a kinetic framework that directly links equilibrium microscale heterogeneity to the macroscopic thermodynamic anomalies of ethanol-water mixtures, and demonstrate that in a simple binary liquid, periodic physical perturbations alone can drive hierarchical ordering without molecular design or external templates. This principle offers a general self-assembly strategy with implications for condensate biology, distilled spirit ageing, and stimuli-responsive soft materials.

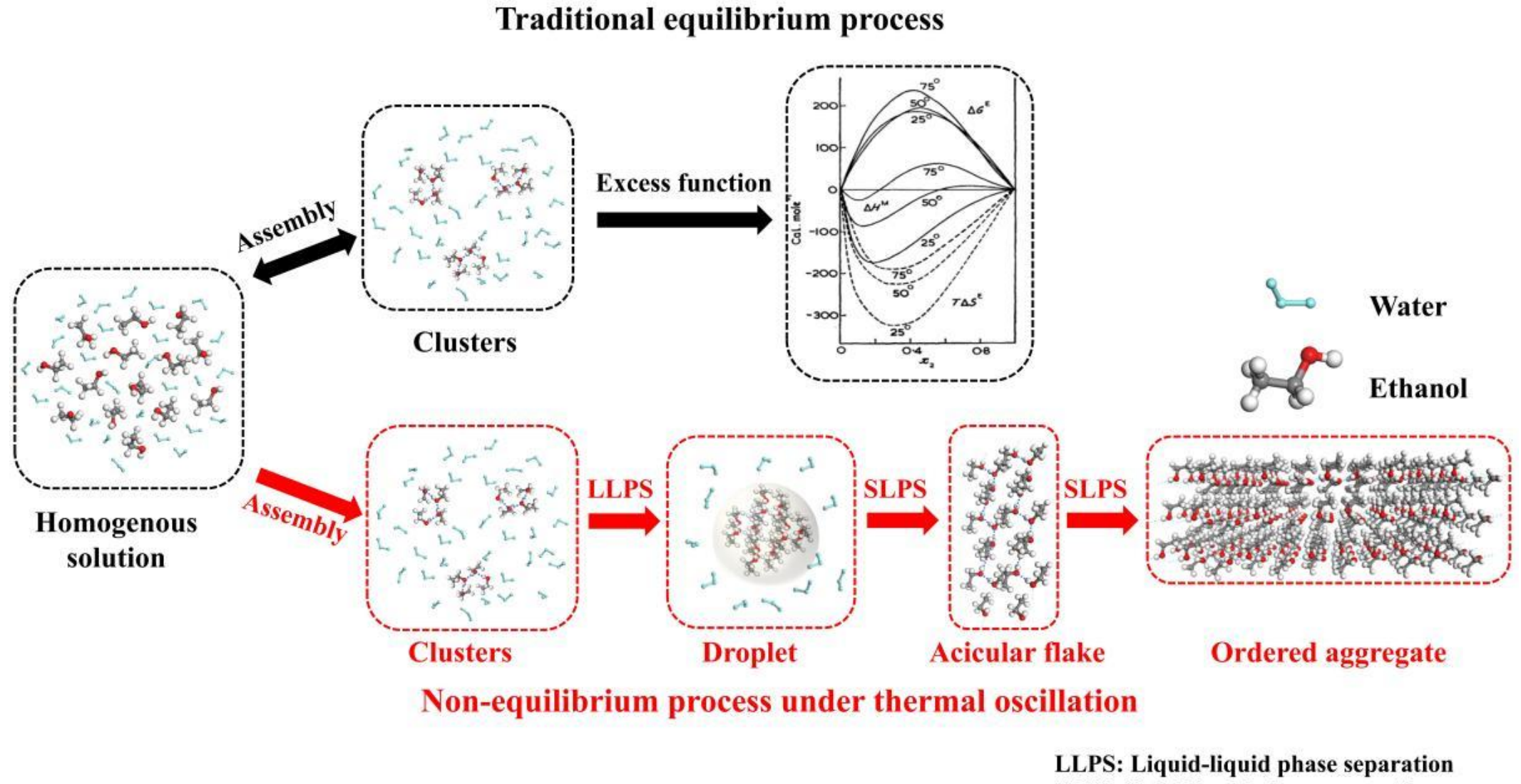

**Fig. 1** Schematic diagram of the hierarchical evolution of ethanol aggregates induced by thermal oscillations in non-equilibrium state

## Results and discussion

### Morphological evolution and phase separation under thermal oscillation

Transient clusters driven by hydrogen-bonding interactions abound in binary ethanol-water mixtures[18, 19, 20]. Their structure and distribution depend strongly on solvent composition, temperature, and external perturbation[21, 22, 23]. Previous studies have shown that these clusters typically remain in a rapid dynamic equilibrium and thus resist preservation on the macroscopic scale. Against this background, we investigated whether periodic thermal oscillation disrupts this dynamic equilibrium and promotes the emergence of higher-order structures.

After 1-cycle thermal oscillation, spherical droplets of relatively uniform size appeared in the ethanol-water solution (Fig. 2a). This observation indicated that the initial clusters rearranged under thermal perturbation and entered an intermediate state governed by liquid-liquid microphase separation. As the number of thermal-oscillation cycles increased, the system developed acicular flakes alongside three-dimensional ordered ethanol aggregates formed through flake stacking. This progression suggested that the droplet intermediate subsequently evolved into ethanol-rich ordered domains during dehydration and structural reorganization. Scanning electron microscopy (SEM) images (Fig. 2b-c) revealed that the acicular flakes were single-layer, flake-like structures with sizes of approximately 10 ~ 20 μm and pronounced anisotropy. Further stacking yielded three-dimensional ordered ethanol aggregates measuring approximately 40 ~ 50 μm. Energy-dispersive X-ray spectroscopy (EDS) analysis (Supplementary Fig. 1) gave a C:O atomic ratio of approximately 2:1 for these structures, consistent with the stoichiometry of ethanol. This result confirmed that the aggregates formed after repeated thermal oscillation were ethanol-dominated.

To track the evolution of these structures, we employed fluorescence microscopy with Rhodamine B/coumarin dual staining (Fig. 2d-h). Rhodamine B preferentially labelled the spherical droplets, whereas coumarin mainly stained the more hydrophobic particulate structures[24, 25]. As the thermal-oscillation cycles increased, the number of droplets rose continuously, while the size and abundance of the flake-like structures increased in parallel, eventually yielding three-dimensional ordered ethanol aggregates at higher cycle numbers. These observations demonstrated that thermal oscillation induced droplet formation through liquid-liquid microphase separation and simultaneously promoted the evolution of ethanol-rich domains from two-dimensional flakes into higher-order assemblies. To further assess the development of hydrophobic microdomains, we performed steady-state fluorescence analysis using the polarity-sensitive probe Nile Red. Previous work has established that the fluorescence intensity of Nile Red increases markedly in low-polarity environments, serving as an effective indicator of local hydrophobic domains[26, 27]. Under 552 nm excitation, the emission intensity at 640 nm rose steadily with increasing thermal-oscillation cycles (Fig. 2i), indicating a progressive enhancement of hydrophobic microdomains. This result was consistent with the dual-staining images and morphological observations, and further demonstrated that ethanol gradually segregated from the original water-dominated network to form more spatially extended, hydrophobic-enriched domains.

Raman spectra (Supplementary Fig. 2) further indicated that these ordered ethanol structures differed from solution-state clusters both in morphology and in their local hydrogen-bonding environment. Compared with anhydrous ethanol at room temperature, low-temperature crystalline ethanol exhibited a new Raman band near 3090 $cm^{-1}$, whereas the ordered ethanol aggregates formed after 10 cycles of thermal oscillation showed a corresponding new band at approximately 3070 $cm^{-1}$. Given that low-temperature ethanol crystals consist of regular O-H···O hydrogen-bonded chains [28], this result suggested that the ordered ethanol aggregates formed under thermal oscillation share certain ordered ethanol-ethanol hydrogen-bonding features with low-temperature ethanol crystals, although their precise structural motif awaits further evidence.

In summary, thermal oscillation drives the ethanol-water system from a state of dynamic clusters into heterogeneous structures with pronounced phase-separation characteristics. The evolution pathway was summarized as follows: formation of water-containing droplets as liquid-liquid microphase-separated intermediates, growth of ethanol flake-like ordered structures during dehydration, and eventual development of three-dimensional ordered ethanol aggregates through flake stacking. These results demonstrated that periodic thermal perturbation effectively disrupted the original rapid dynamic equilibrium and guided the ethanol–water solution through a hierarchical evolution-from liquid-liquid microphase separation to the establishment of ethanol ordered aggregates with solid-like characteristics.

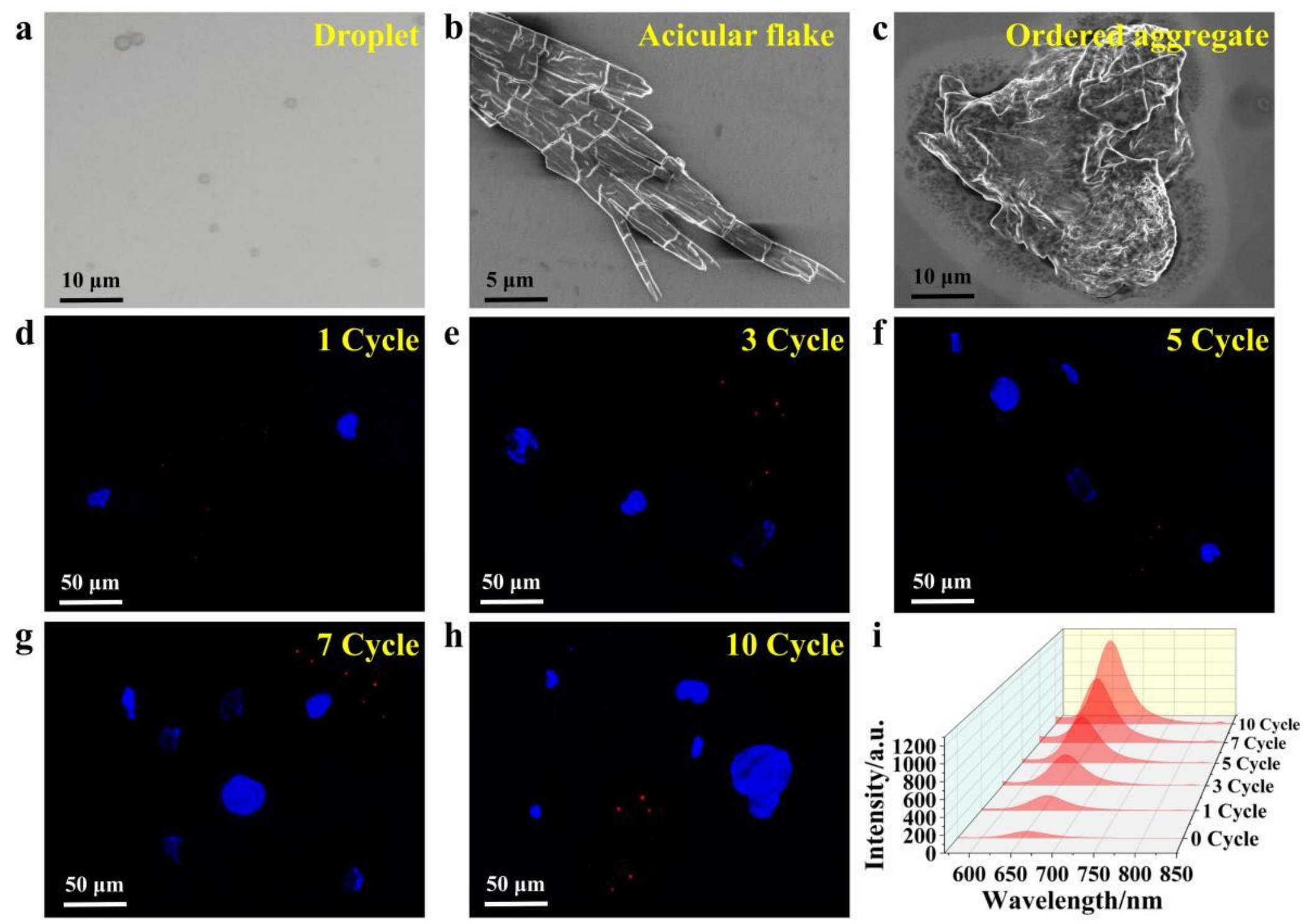

**Fig. 2** Morphological evolution and fluorescence characterization. (a) Optical microscopy image of ethanol-water droplets; (b, c) SEM images of ethanol-water acicular flake and ordered ethanol aggregate. Dual-dye fluorescence microscopy images of ethanol-water solutions under different thermal oscillation cycles using coumarin (excitation wavelength λex=330 nm, blue fluorescence) and Rhodamine B (λex=525 nm, red fluorescence): (d) 1-cycle; (e) 3-cycle; (f) 5-cycle; (g) 7-cycle; (h) 10-cycle. (i) Steady-state fluorescence emission spectra using Nile Red as probe (λex=552 nm).

**Thermal stability and hierarchical organization of ordered ethanol aggregates**

Differential scanning calorimetry (DSC) evaluated the thermal behavior of the aggregated structures formed in ethanol-water solutions after different thermal-oscillation cycles[29, 30]. As shown in Fig. 3a, the untreated control sample exhibited only a broad endothermic peak at 33.9 °C, reflecting a diffuse endothermic process in the ethanol-water mixture. After 1-cycle thermal oscillation, a sharp endothermic peak emerged at 22.87 °C, indicating that thermal perturbation had already induced a new aggregated state with a distinct thermal response. With increasing thermal-oscillation cycles, this peak shifted progressively toward higher temperatures, reaching 24.76, 28.18 and 39.76 °C after 3, 5 and 7 cycles, respectively. This trend indicated a gradual increase in the thermal stability of the ethanol-related aggregates. After 10 cycles, the heat-flow curve displayed three resolvable endothermic peaks at 14.21, 21.86 and 40.6 °C, demonstrating the coexistence of at least three aggregated components with different thermal stabilities. In conjunction with the morphological results described above, these three thermal events were assigned to cluster, droplet and flake-stacked ordered ethanol aggregates, respectively. Thus, thermal-oscillation treatment increased the stability of a single aggregated state and established a multilevel population of aggregates that coexisted after extensive cycling.

The enthalpy changes (ΔH) obtained by integrating the endothermic peaks are summarized in Fig. 3b. With increasing thermal-oscillation cycles, ΔH increased from -6.71 kJ/mol for the untreated control to -83.11 kJ/mol after 10 cycles, with values of -6.71, -19.86, -25.51, -36.39, -78.87 and -83.11 kJ $mol^{-1}$ after 0, 1, 3, 5, 7 and 10 cycles, respectively. The steady increase in the absolute value of ΔH indicated that the fraction of ordered aggregated structures continuously increased and that the system progressively evolved from a weakly associated initial state to a highly cooperative aggregated state[31]. Based on the DSC results together with the morphological assignments, the apparent free energies associated with cluster, droplet and flake were further estimated to be 24.55, 42.64 and 83.13 kJ/mol, respectively (Fig. 3c). Although these values serve as relative indicators of stability differences among the aggregated states rather than rigorous thermodynamic functions, their trend remained fully consistent with the degree of structural order and thermal stability: the energetic scale associated with formation and dissociation increased continuously from cluster to droplet and then to flake.

Due to the temperatures derived from DSC only represented apparent thermal responses under programmed heating conditions[32, 33], the actual dissociation behavior of representative aggregates was further examined by in situ temperature-dependent Raman spectroscopy combined with optical microscopy. The $CH_3$/$CH_2$ stretching band near 2900 $cm^{-1}$ served as the spectroscopic marker for the ethanol aggregates[34, 35], and the dissociation temperature was determined from abrupt intensity loss together with the corresponding morphological change. As shown in Fig. 3d, when the laser focused on a droplet, the ethanol-associated Raman signal rapidly decayed to the noise level once the temperature reached 115 °C, while the corresponding droplet disappeared completely from the microscopic field of view (Supplementary Fig. 3a), indicating complete disintegration of the droplet near this temperature. In contrast, the characteristic Raman signal of the flake-stacked ordered ethanol aggregate did not disappear until 165 °C (Fig. 3e), and the corresponding flake-like morphology was simultaneously disrupted (Supplementary Fig. 3b). These observations demonstrated that the flake possessed substantially higher thermal stability than the droplet, consistent with the assignment of the higher-temperature endothermic peak in the DSC traces to a more ordered aggregated structure.

To further resolve the hierarchical composition of these aggregates and their thermal dissociation behavior at the molecular level, we performed GC-MS on 10-cycle of thermal oscillation sample. In this work, electron-impact ionization was used. Consequently, the ion signals detected during programmed heating primarily reflected dissociation and fragmentation of the aggregates during vaporization and electron bombardment as opposed to intact cluster ions[36, 37, 38]. The terms “tetramer”, “decamer” and “icosamer” used here denote hierarchical assignments based on characteristic fragment combinations[39], their temperature windows of appearance and consistency with the in situ Raman and DSC results. These assignments derive from fragment evidence rather than from direct observation of intact oligomeric ions. The complete list of characteristic ions, corresponding temperature windows and tentative assignments appeared in Tables S1-S2 and Supplementary Fig. 4. As shown in Fig. 3f, the sample exhibited pronounced temperature-dependent fragmentation during programmed heating, indicating the presence of multiple ethanol-associated components with different thermal stabilities after thermal oscillation. The fragments appearing in the lower-temperature range corresponded to weakly bound early-stage aggregates[39]. In the window shown in Supplementary Fig. 4a, the ions at m/z 142.077 and 96.093 observed at 95 °C were assigned to cluster-related low-level species. In the window shown in Supplementary Fig. 4b, the ion at m/z 134.072 observed at 125 °C fell within the dissociation range of the droplet and therefore reflected a more weakly associated droplet-related aggregate. By contrast, the signals shown in Fig.s 3g and 3h corresponded to more stable components appearing at higher temperatures. The ion at m/z 184.087 observed at 185 °C was assigned to a tetramer-level skeletal ion, indicating that the flake-related aggregate still released a stable low-order structural unit under high-temperature conditions. In the retention-time window around 20.8 s, the ion at m/z 110.108 observed at 185 °C together with the ion at m/z 149.132 observed at 125 °C constituted a decamer-level fragment set, demonstrating that higher-level ethanol aggregates generated both high-temperature fragments and lower-mass stable core ions during dissociation. Furthermore, in the window shown in Fig. 3i, the ion at m/z 149.023 observed at 125 °C was assigned to an icosamer-level fragment set, indicating that higher-order aggregates at longer retention times similarly relaxed to a related low-order core during fragmentation. A notable feature was that m/z 149.0x repeatedly appeared in different retention-time windows and therefore represented one of the most characteristic common core fragments in this system. Given its consistency across the dissociation of different hierarchical aggregates, this signal was reasonably assigned to a stable core ion derived from dehydration of a tetrameric ethanol skeleton, suggesting that tetramer-level low-order units served as important shared structural modules among different higher-order aggregates.

Overall, above results demonstrated that thermal oscillation drove the ethanol-water system through a hierarchical evolutionary pathway. The aggregates formed varied in thermal stability and structural order, progressing from loosely bound clusters and droplets to highly stable, flake-stacked ordered assemblies. The recurring tetramer-level core fragments across higher-order species further suggested that these flake-stacked structures arose from the ordered assembly of specific low-order modules rather than from random molecular aggregation.

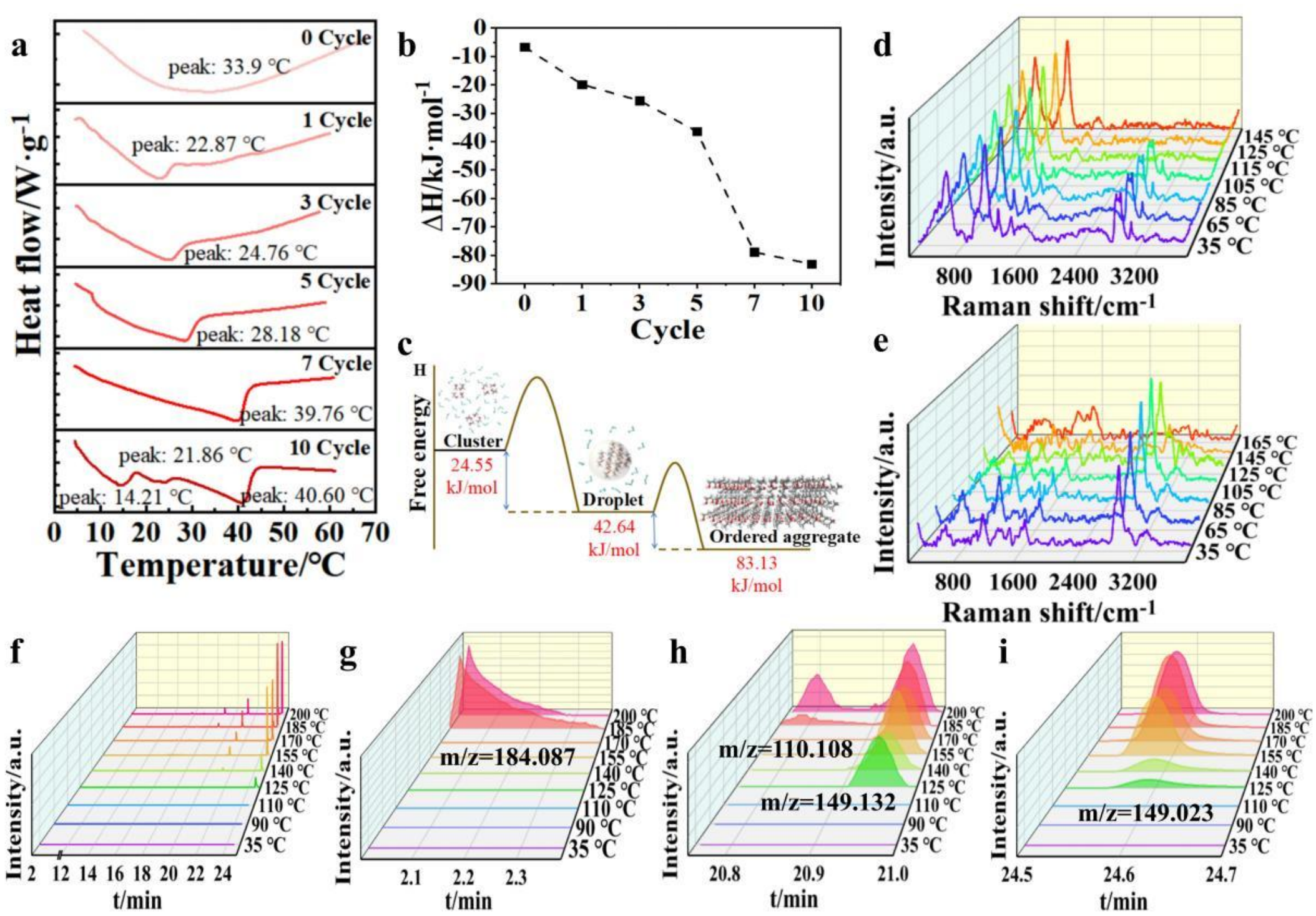

**Fig. 3** Multiscale characterization of thermal stability and dissociation behavior. (a) DSC heat flow curves of ethanol-water solutions under different thermal oscillation cycles; (b) Enthalpy change (ΔH) derived from DSC integration as a function of thermal oscillation cycles; (c) Schematic illustration of the free energy evolution associated with ethanol phase transition; (d-e) In situ temperature-dependent Raman spectra of ethanol-water droplets and flake structures; (f) Raw mass spectra recorded during temperature ramping; (g-i) Characteristic fragment signals corresponding to ethanol tetramer-level, decamer-level, and icosamer-level, appearing at approximately 2.0 s, 20.8 s, and 24.5 s, respectively.

## Macroscopic consequences of ethanol aggregation

The morphological observations, thermal analyses and variable-temperature mass spectrometry described above showed that thermal oscillation drove the ethanol-water system toward more ordered and more stable aggregated structures. To determine whether this structural reorganization transmitted to the macroscopic scale, we measured the viscosity of ethanol-water solutions subjected to different thermal-oscillation cycles using an Ubbelohde viscometer. As shown in Fig. 4a, the viscosity increased continuously with the thermal-oscillation cycles and was 15.71% higher after 10 cycles than in the untreated control. This change indicated that the ethanol-rich structures induced by thermal oscillation were no longer confined to local transient fluctuations but instead had sufficiently developed to alter the overall flow behavior of the system[40, 41].

The viscosity increase alone provided insufficient evidence to establish whether ethanol molecules had indeed transitioned from a freely dispersed state into an aggregated state. To address this issue, we further quantified the free ethanol content through the alcohol dehydrogenase (ADH) reaction. ADH effectively catalyzed only ethanol molecules that were freely accessible to the enzyme active site[42]. The extent of reaction therefore served as a functional measure that distinguished free ethanol from aggregated ethanol[35, 43]. During the reaction, oxidation of ethanol was accompanied by reduction of $NAD^+$ to NADH, and the amount of NADH formed was quantified by fluorescence intensity. As shown in Fig. 4b, the NADH fluorescence decreased steadily with increasing numbers of thermal-oscillation cycles, indicating that the fraction of ethanol available for enzymatic consumption progressively declined. A calibration relationship between NADH fluorescence intensity and ethanol concentration was established using ethanol-water solutions of 35, 40, 45 and 50% (v/v) ethanol (Fig. 4c). On the basis of this calibration, the actual free ethanol contents of the initial 50% (v/v) ethanol-water system after 1, 3, 5, 7 and 10 cycles of thermal oscillation were determined to be 47.13, 43.89, 41.86, 40.35 and 39.19%, respectively. These values showed that, although the total ethanol concentration remained unchanged, thermal oscillation progressively decreased the proportion of ethanol that remained in the free state and participated in the enzymatic reaction.

Thus, these results demonstrated that thermal oscillation substantially altered the mode of ethanol incorporation in the ethanol-water system. As the thermal-oscillation cycles increased, the viscosity rose continuously whereas the free ethanol content decreased steadily, indicating a gradual transition of ethanol molecules from a freely dispersed state to an aggregated state. From the perspectives of macroscopic properties and enzymatic response, these observations further confirmed that thermal oscillation effectively promoted the formation and accumulation of ethanol aggregates.

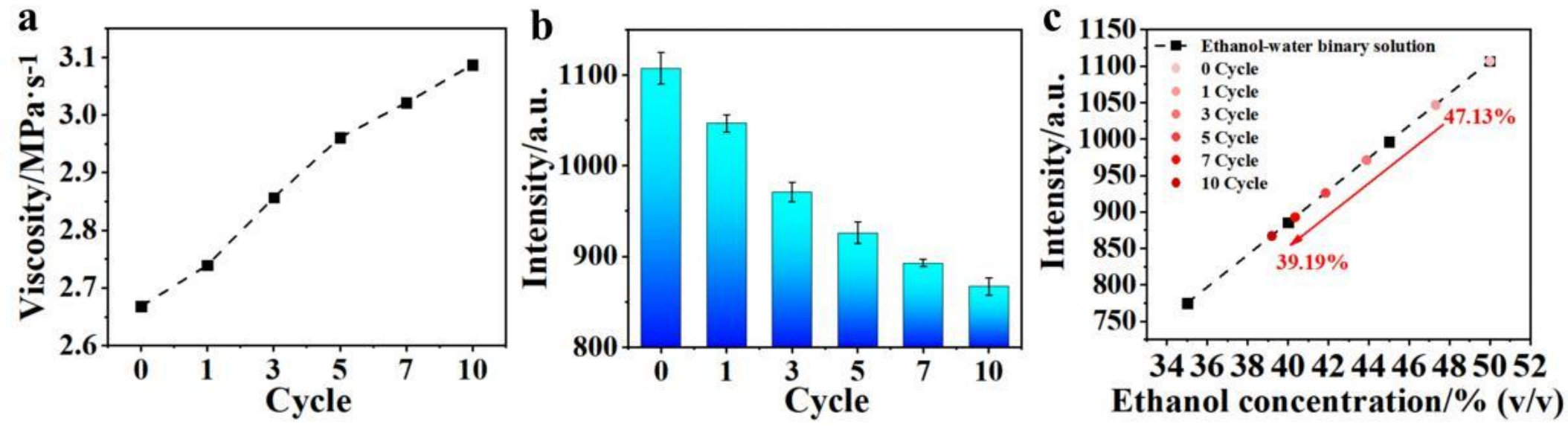

**Fig. 4** Effects of thermal oscillation on macroscopic properties of ethanol-water solutions under different cycles of thermal oscillation treatment. (a) Viscosity values of ethanol-water solutions; (b) Fluorescence intensity of NADH in ethanol-water solutions; (c) Function curves of fluorescence intensity and alcohol content of NADH in ethanol-water binary solutions with different concentrations.

**Formation kinetics of ordered ethanol aggregate**

To elucidate the rearrangement of the hydrogen-bond network in ethanol–water mixtures during thermal oscillation and the formation kinetics of ordered ethanol aggregates, we recorded steady-state fluorescence spectra under 240 nm excitation for samples subjected to different thermal-oscillation cycles (Fig. 5a). Gaussian deconvolution resolved the emission profiles into three characteristic bands at 308, 330, and 373 nm, corresponding to ethanol-water associated clusters in different hydrogen-bonding environments, which were assigned to local structures dominated by ethanol-rich $(H2O)(EtOH)_m$, water-ethanol co-associated $(H2O)_n(EtOH)_m$, and water-dominated $(H2O)_n(EtOH)$ motifs, respectively[44]. With increasing thermal-oscillation cycles, all three emission bands increased progressively and exhibited an overall blue shift, indicating systematic changes in the polarity and configurational distribution of the local microenvironments[45, 46]. In molecularly associated systems, enhanced fluorescence generally reflects a higher degree of local structural ordering together with suppression of intra- and intermolecular nonradiative decay[47, 48]. These observations therefore indicated that thermal oscillation promoted the evolution of the system from a relatively disordered freely dispersed state to a more ordered associated state[49]. A notable feature was the particularly pronounced increase of the 330 nm band, indicating preferential accumulation of the water–ethanol co-associated intermediate. This behavior suggested that thermal oscillation first weakened the solvation constraint imposed by the original water-dominated hydrogen-bond network, thereby releasing a fraction of ethanol into a locally enriched environment characterized by ethanol-ethanol interactions. To further verify the spectroscopic signature of the highly stable aggregates, we naturally dried samples subjected to different thermal-oscillation cycles in order to preserve, as far as possible, the stable ethanol-rich aggregates generated during thermal oscillation. We then measured their steady-state fluorescence spectra under the same excitation conditions (Fig. 5b). Relative to the solution-state samples, the dried samples displayed an emission band that was markedly red-shifted to approximately 450 nm, and its intensity increased continuously with increasing numbers of thermal-oscillation cycles. This emission feature was clearly distinct from the 308 ~ 373 nm region observed for the solution-state samples, indicating that the stable ethanol aggregates formed after thermal oscillation retained an independent emissive microenvironment even after removal of the bulk solvent. Thus, thermal oscillation generated not ordinary transient and reversible solution clusters but highly stable ordered ethanol aggregates that remained partially preserved during subsequent drying[50, 51].

To determine the sequence linking disruption of the water network with formation of ethanol-rich aggregates, we further analyzed the steady-state fluorescence data by two-dimensional correlation spectroscopy (2D-COS)[52]. In the synchronous spectrum (Fig. 5c), a pronounced positive cross peak appeared between 308 and 373 nm, indicating that attenuation of the water-dominated structure and enhancement of the ethanol-rich structure proceeded cooperatively on the overall scale. The asynchronous spectrum (Fig. 5d) provided additional information on the sequence of these changes. According to Noda’s rules[53], the evolution order of the spectral bands was established as 373 nm → 330 nm → 308 nm, indicating that the water-dominated local environment responded first, followed by the water–ethanol co-associated intermediate, and finally by the establishment of the ethanol-rich aggregate. These results showed that the different structures changed sequentially along a well-defined pathway rather than simultaneously and randomly.

This kinetic pathway was subsequently corroborated by $^1H$ NMR. Hydrogen-bond network in alcohol-water mixtures determines the local coordination environment and molecular organization, variations in hydroxyl proton chemical shifts and integrals directly reflect hydrogen-bond reorganization[54, 55, 56]. We measured samples subjected to different numbers of thermal-oscillation cycles on a 600 MHz $^1H$ NMR spectrometer (Bruker Avance III HD). The spectra shown in Supplementary Fig. 5 were obtained after using the $CH_2$ resonance of sodium propionate as an internal standard and normalizing its chemical shift to 2.200 ppm. The internal-standard signal retained constant chemical shift and integral intensity before and after thermal oscillation, demonstrating good comparability among the samples. By contrast, the ethanol $CH_3$ (1.10 ppm) and $CH_2$ (3.560 ppm) signals decreased continuously with increasing numbers of thermal-oscillation cycles and were reduced to only very weak responses after 10 cycles, indicating that a large fraction of ethanol molecules, particularly their hydrophobic ethyl moieties, had left the NMR-visible bulk liquid phase and entered aggregated states that conventional liquid-state $^1H$ NMR failed to detect effectively.

In Fig. 5e, the ethanol hydroxyl resonance (E-OH, $P_C$) was magnified 10-fold for comparison. Both $P_C$ and the water hydroxyl resonance (W-OH, $P_T$) shifted downfield with increasing numbers of thermal-oscillation cycles[21]. Specifically, PC moved from 4.768 to 4.811 ppm, whereas $P_T$ shifted from 5.354 to 5.401 ppm (Fig. 5f). The simultaneous downfield shifts indicated strengthening of the average hydrogen-bond interaction in the residual observable phase after thermal oscillation, showing that the system underwent hydrophobic segregation and experienced reorganization of the hydrogen-bond network. Further analysis based on MestReNova integration of the four characteristic peaks, with the $P_C$ integral normalized to 1.000, yielded the values shown in Fig. 5g. After 10 cycles of thermal oscillation, the relative proton fractions assigned to $P_C$, $P_T$, $CH_2$, and $CH_3$ changed from 7.51 to 6.96%, 64.4 to 90.59%, 11.27 to 1.04%, and 16.76 to 1.39%, respectively. The marked decreases in the $CH_2$ and $CH_3$ fractions indicated that a large proportion of ethanol molecules no longer remained in the bulk solution as freely dispersed species. The corresponding increase in the $P_T$ fraction showed that the NMR-observable portion progressively became a continuous phase dominated by water together with a small amount of residual hydroxyl-containing species. Thus, thermal oscillation produced not uniform ordering throughout the entire system but rather a pronounced spatial redistribution of ethanol, with one fraction remaining in the bulk phase and another entering aggregated states that conventional liquid-state NMR failed to detect.

To further resolve the dynamics of this observable phase, we performed relaxation experiments in a 50% (v/v) deuterium oxide-ethanol-$D_6$ system to avoid interference from ordinary water proton signals during parameter fitting. We measured T1 values by the inversion-recovery method and determined T2 values with a CPMG pulse sequence[57, 58]. Relative to the untreated sample, after 10 cycles of thermal oscillation the T2 values of W-OH, E-OH, $CH_2$ and $CH_3$ increased by 1.27, 0.83, 4.94 and 1.44%, respectively (Fig. 5h). Although the magnitude of these changes was limited, their consistent direction indicated that the local environment of the NMR-observable phase became more homogeneous after thermal oscillation. The most reasonable interpretation was that a large amount of ethanol had migrated out of the bulk phase into larger and more motionally restricted ordered aggregates, whereas these aggregates contributed little to conventional liquid-state NMR because of very fast relaxation and substantial line broadening[59]. The measured T2 values therefore reflected the dynamics of the residual mobile phase. As the concentration of free ethanol decreased, intermolecular interactions in the bulk phase weakened and the local environment became more homogeneous, resulting in the overall increase in T2.

The T1/T2 ratio provided further information on the degree of motional isotropy for the observable groups[60, 61], as shown in Fig. 5i. After thermal oscillation, the T1/T2 ratio of E-OH increased by 1.04%, whereas those of W-OH, $CH_2$ and $CH_3$ decreased by 1.85, 7.88 and 4.12%, respectively. Given that all T1/T2 values remained generally close to unity, the observable species were inferred to remain in the fast-motion limit and no directly distinguishable slow condensed phase was detected[62]. For this reason, the relaxation measurements probed the dynamical reorganization of the residual bulk phase after aggregate formation, whereas the large aggregates themselves remained outside the detection window of these measurements. Following migration of ethanol out of the bulk phase, the hydrogen-bonding environments of the hydroxyl-containing species and water were redistributed, the effective concentration of alkyl groups decreased substantially, and the observable phase consequently exhibited a more dilute and more homogeneous fast-dynamic character.

Overall, the steady-state fluorescence, 2D-COS and NMR results consistently showed that thermal oscillation induced rearrangement of a mixed state and simultaneously drove a non-equilibrium self-assembly process with a well-defined sequence. This process began with perturbation and relaxation of the water-dominated hydrogen-bond network, proceeded through a ethanol-water co-associated intermediate stage, and ultimately led to sustained enrichment of desolvated ethanol and formation of highly stable ordered ethanol aggregates. These observations established the role of thermal oscillation in driving the formation and stabilization of ordered ethanol aggregates from the perspectives of kinetic pathway, molecular redistribution and local intermolecular interactions.

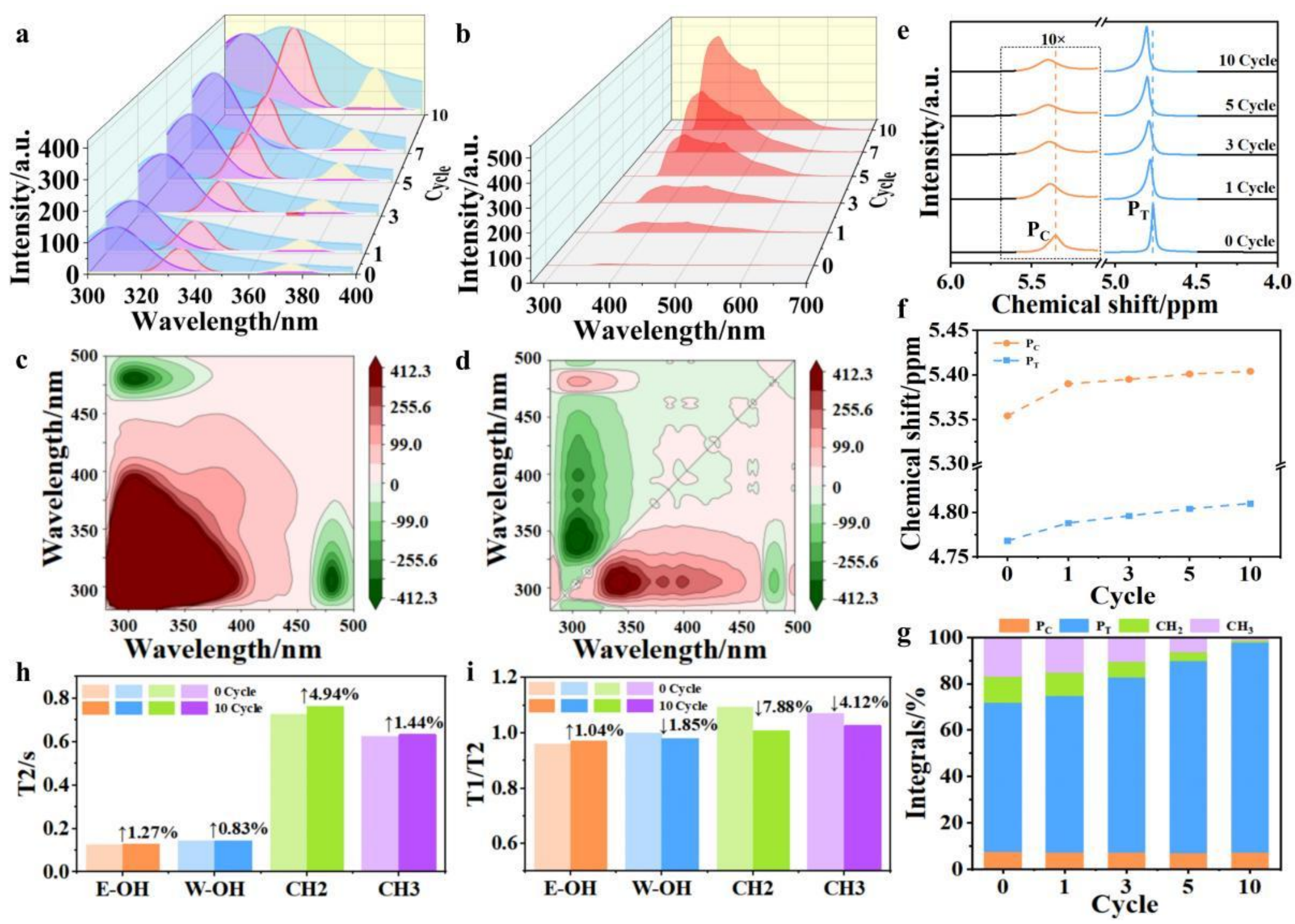

**Fig. 5** Spectroscopic and NMR characterization of microstructural evolution in the ethanol-water solution under different cycles of thermal oscillation treatment. (a) Steady-state fluorescence spectra with Gaussian peak deconvolution (λex=240 nm); (b) Steady-state fluorescence spectra of the air-dried ordered ethanol aggregates samples (λex=240 nm); (c-d) Asynchronous and synchronous two-dimensional correlation spectra; (e) $^1$H NMR spectra of ethanol-water solutions with the hydroxyl peaks of water ($P_T$) and ethanol ($P_C$, magnified to 10 times); (f) Chemical shift variations of $P_C$ and $P_T$; (g) Relative proportions of $P_T$, $P_C$, methylene ($CH_2$), and methyl ($CH_3$) groups derived from peak integration; (h) Transverse relaxation times (T2) of 50% (v/v) Deuterium oxide-ethanol-D6 solution under 0 and 10 cycles of thermal oscillation treatment; (i) Corresponding T1/T2 ratio.

## Conclusions

This work demonstrated that periodic thermal oscillations propelled the ethanol-water system, a classic binary mixture, from equilibrium local fluctuations into a complete, traceable non-equilibrium hierarchical assembly pathway. Rather than merely accelerating relaxation, thermal oscillation guided the system along a specific kinetic route from ethanol clusters, through a transient ethanol-water coexisting intermediate, to stable ordered ethanol aggregates. Our findings revealed that the longstanding thermodynamic anomalies of ethanol-water mixtures were not equilibrium properties, but rather the calorimetric fingerprint of an arrested phase transition. This insight transcended the static framework of conventional solution theory and offered a kinetic perspective on the structural origin of solution non-ideality, where volume contraction, exothermic mixing, and the large negative excess entropy were natural consequences of hierarchical assembly and cooperative hydrogen-bond network reorganization under non-equilibrium driving.

More broadly, this work established that in the simplest of binary liquids, macroscopic order could arise solely from periodic physical perturbations, without sophisticated molecular design or external templates. This principle, that periodic energy input guided simple systems toward complex order, provided a new paradigm for soft matter self-assembly, with broad implications for understanding dynamic condensate assembly in living systems, modulating flavour evolution during distilled spirit ageing, and designing novel stimuli-responsive materials.

## Methods

**Chemicals, sample preparation and thermal oscillation treatment**

A 50% (v/v) ethanol-water solution was prepared using deionized water and anhydrous ethanol. Portions of the solution were sealed in polytetrafluoroethylene-lined stainless-steel autoclaves and subjected to 0, 1, 3, 5, 7, and 10 programmed thermal-oscillation cycles. Each cycle consisted of heating to 80 °C at 5 °C/min, holding at 80 °C for 30 min, and cooling to 25 °C at 5 °C/min. Complete lists of reagents, purities, and suppliers are provided in the Supplementary Information.

**Morphological characterization**

Droplets of the thermally oscillated solution were dried on glass slides or silicon wafers and examined with an optical microscope (BX53, Olympus) and a scanning electron microscope (SU8230, Hitachi) equipped for energy-dispersive X-ray spectroscopy (EDS).

**Raman spectroscopy**

Raman spectra were collected on a confocal Raman spectrometer (XploRA Plus, HORIBA) using 532 nm excitation. Liquid samples, air-dried ordered aggregates, and anhydrous ethanol crystallized at -140 °C were measured to probe molecular organization.

**Fluorescence characterization**

Samples were labeled with coumarin and Rhodamine B (6 nM each) and imaged on a Nikon C1 confocal microscope with 330 nm and 525 nm excitation, respectively. Steady-state fluorescence spectra of both liquid and dried aggregate samples were recorded on an F-7000 spectrophotometer (Hitachi) with an excitation wavelength of 240 nm. The hydrophobic probe Nile Red (60 nmol/L final concentration) was added to the solutions and excited at 552 nm. Two-dimensional correlation analysis was applied to the series of fluorescence spectra.

**Thermal analysis**

Differential scanning calorimetry was performed on a Q20 calorimeter (TA Instruments) from 0 to 65 °C at 1 °C/min. In-situ Raman spectra during heating were acquired on a sample dried on a silicon wafer using a heating stage.

**GC-MS analysis**

Ordered aggregates from the 10-cycle sample were collected onto an SPME fiber and introduced into a gas chromatograph-mass spectrometer (TRACE 1310 GC coupled to a Q Exactive GC Orbitrap, Thermo Scientific) via a programmed-temperature vaporization (PTV) inlet. Gradient thermal desorption from 50 to 320 °C was performed in 5 °C steps, with electron ionization (70 eV) and a mass range of 35 ~ 400 m/z. Data were processed with Compound Discoverer 3.3 and identified by matching retention indices and fragmentation patterns against the NIST 2.3 library.

**Viscosity and ADH activity assays**

Absolute viscosities were measured at 20 °C with an Ubbelohde viscometer. Alcohol dehydrogenase activity was assessed by monitoring NADH production at 340 nm (ELISA plate reader, BioTek ELx808) at 25 °C and pH 8.0.

**NMR spectroscopy**

$^{1}H$ NMR spectra and relaxation measurements were performed at 600 MHz (Bruker Avance III HD) at 25 °C. Samples were diluted in 50% (v/v) $D_2O$/ethanol-$D_6$, with sodium propionate as an external reference. T1 values were determined by the inversion-recovery method, and T2 values by the CPMG sequence (echo spacing $\tau$ = 400 μs). Detailed parameters, fitting procedures, and processed spectra are given in the Supplementary Information.

## Data availability

All data are included within the Article and its Supplementary Information.

## Acknowledgements

We acknowledge Xu Zhang and Tao Huang (State Key Laboratory of Magnetic Resonance and Atomic Molecular Physics, National Center for Magnetic Resonance in Wuhan, Innovation Academy for Precision Measurement Science and Technology, Chinese Academy of Sciences) for help with Nuclear Magnetic Resonance experiments.

## Author information

### Authors and Affiliations

School of Civil Engineering Architecture and Environment, Hubei University of Technology, Wuhan, China (Xinyue Jiang)

School of Life and Health Sciences, Hubei University of Technology, Wuhan, China (Yating Shang, Yuqun Xie)

State Key Laboratory of Advanced Technology for Materials Synthesis and Processing, Wuhan University of Technology, Wuhan, China (Jianhui Li, Zhaoyong Zou)

The Analysis and Testing Center of Institute of Hydrobiology, Chinese Academy of Sciences, Wuhan, China (Yanxia Zuo)

**Contributions**

X.J., Z.Z., and Y.X. conceived the project and designed the experiments. X.J. and Y.S. performed morphological characterization, fluorescence characterization, thermal analysis, viscosity and ADH activity assays. X.J. and J.L. performed SEM and Raman spectroscopy. Y.Z. performed GC-MS analysis. Z.Z. provided technical assistance. X.J. and Y.X. wrote the original manuscript. All authors discussed the results and revised the study. Z.Z. and Y.X. supervised the research.

**Corresponding author**

Correspondence to Yuqun Xie.

## Ethics declarations

Competing interests

The authors declare no competing interests.

## Supplementary information

Supplementary Methods, Tables 1-2, Figures 1-7.